\documentclass[conference]{IEEEtran}
\usepackage{amssymb}
\usepackage[cmex10]{amsmath}
\usepackage{stfloats}
\usepackage{graphicx}
\usepackage{subfigure}
\usepackage{tabularx}
\usepackage{epsfig,epsf,color,balance,cite}
\usepackage{verbatim}

\usepackage{color}

\definecolor{myc1}{rgb}{0,0,0}

\newtheorem{lemma}{Lemma}
\usepackage{algorithm}
\usepackage{algpseudocode}
\usepackage{amsmath}

\begin{document}

\title{Joint Time-Frequency Splitting for Multiuser SWIPT OFDM Networks}

\author{
\IEEEauthorblockN{Zhaohui Yang \IEEEauthorrefmark{1},
                  Wei Xu \IEEEauthorrefmark{2},
                  and Mohammad Shikh-Bahaei\IEEEauthorrefmark{1},
                  }
\IEEEauthorblockA{\IEEEauthorrefmark{1}Centre for Telecommunications Research, Department of Informatics, King’s College London, WC2B 4BG, UK}
\IEEEauthorblockA{\IEEEauthorrefmark{2}National Mobile Communications Research Laboratory, Southeast University, Nanjing 211111, China}
\IEEEauthorblockA{E-mail: yang.zhaohui@kcl.ac.uk, w.xu@seu.edu.cn, m.sbahaei@kcl.ac.uk }}

\IEEEtitleabstractindextext{
\begin{abstract}

In this paper, we propose a joint time-frequency splitting (TFS) strategy for a multiuser orthogonal frequency division multiplexing (OFDM) system with simultaneous wireless information and power transfer (SWIPT). In TFS, the time sharing factors for each user on different subcarriers are optimized via maximizing the sum rate of users with both information and energy quality of service (QoS) constraints. Though the original problem is nonconvex, we first transform it into an equivalent convex problem through an appropriate variable transformation. Then, we present an iterative algorithm based on semi closed form with low complexity. Numerical results show that the proposed TFS outperforms the conventional time-sharing and subcarrier-separation strategies.


\end{abstract}

}

\maketitle
\IEEEdisplaynontitleabstractindextext
\IEEEpeerreviewmaketitle

\section{Introduction}
\label{section1}

Recently, simultaneous wireless information and power transfer (SWIPT) has attracted much attention in academia \cite{4595260}.
Using this technology, users can simultaneously receive information and harvest energy in a wireless way.
This distinguished advantage makes SWIPT a promising technology especially for wireless communications in extreme environment \cite{7317504}, where energy charging is tough work.
SWIPT can be utilized in many wireless systems, such as multiple-input multiple-output (MIMO) \cite{Zhang2011MIMO}, 
cooperative relay networks \cite{6779694}, 
and orthogonal frequency division multiplexing (OFDM) \cite{zhou2014wireless}. 
In the area of Internet of Things, SWIPT can be of fundamental importance for energy supply and information exchange with numerous ultra-low power sensors \cite{miorandi2012internet,yang2018energy,yang2017energy,7506268,yang2018optimal}. 

OFDM is a well designed technology for high-rate wireless communication \cite{yang2017joint}.
However, the performance of the system is {\color{myc1}{usually}} limited by available energy of devices.
{\color{myc1}{To further improve the performance}}, SWIPT has been applied in OFDM systems \cite{6623062,zhou2014wireless,7154495,7986955,7275174,7378525}.
There are mainly two SWIPT strategies for OFDM applications, namely time sharing (TS) \cite{zhou2014wireless} and power splitting (PS) \cite{6623062}.
With the TS strategy,
the received signal is either processed for energy harvesting or information decoding at a single time-duration \cite{zhou2014wireless}.
With the PS strategy, the received signal is split into two parts by a power splitter, with one part for energy receiver and the other for information receiver simultaneously.
In OFDM systems with PS,
specific impacts of resource allocation on system throughput \cite{7154495}, max-min fairness \cite{7986955} and physical-layer security \cite{7275174,7378525,8485781} were investigated.
Considering both TS and PS, a hybrid TS/PS scheme was proposed in \cite{7792203}  for OFDM systems to securely transmit data and transfer energy to a legitimate receiving node.

Different from TS and PS, a subcarrier separation (SS) strategy was recently proposed in \cite{yinresource,7859341}.
Information and power were transferred separately on different subcarriers in the SS strategy \cite{yinresource}, which was proven to outperform TS when more power is required to be transferred.
To maximize the harvested energy, a joint subcarrier and power allocation problem was formulated for PS in \cite{7859341} subject to the information decoding constraint.
However, the performance of SS strategy can be further improved by jointly considering time and frequency splitting.

In order to achieve benefits from both time and frequency domains,
we propose a joint time-frequency splitting (TFS) strategy, where the time sharing factors for each user among different subcarriers are adaptively optimized.
{\color{myc1}{In this paper, we investigate the optimal resource allocation and power control for SWIPT in a downlink multiuser OFDM system with the TFS strategy,
which is different from our previous work in \cite{LiSum2017} focusing on the sum rate maximization problem in a downlink VLC system with SWIPT.}}

We {\color{myc1}{aim at maximizing}} the sum rate of all users subject to both minimal rate requirements and minimal harvested power constraints.
Due to different time factors for different subcarriers, a user can simultaneously receive information and harvest power separately on various subcarriers, which can be implemented by using band-pass filter based OFDM receiver \cite{4623916}.
The contributions of this paper are summarized as follows:
\begin{enumerate}
  \item The sum rate maximization problem for the proposed TFS strategy is formulated. Although the original problem is nonconvex, the problem can be equivalently transformed into a convex one through an appropriate variable transformation.
  \item  By introducing a small positive constant to modify the objective function of the original problem, we successfully obtain the semi closed-form expressions of primal variables.
We show that the gap of the optimal value between the modified problem and the original problem approaches zero as the introduced small positive constant approaches zero.
\item  We propose a novel iterative resource allocation and power control algorithm based on semi closed form to obtain the optimal solution.
Numerical results {\color{myc1}{verify}} that the proposed TFS strategy outperforms the existing TS and SS strategies under various scenarios.
\end{enumerate}

This paper is organized as follows.
In Section $\text{\uppercase\expandafter{\romannumeral2}}$, we introduce the system model and provide the formulation of sum transmission rate maximization problem.
Section $\text{\uppercase\expandafter{\romannumeral3}}$ provides
the optimal condition, and proposes a semi closed form based efficient solution.
The numerical results are displayed in Section $\text{\uppercase\expandafter{\romannumeral4}}$
and conclusions are finally drawn in Section $\text{\uppercase\expandafter{\romannumeral5}}$.

\section{System Model and Problem Formulation}

\label{section2}
Consider a downlink OFDM-based single-cell network over Rayleigh flat fading channels.
In this network, there are $K$ users and $N$ subcarriers, denoted by sets $\mathcal K=\{1, 2, \cdots, K\}$ and $\mathcal N=\{1, 2, \cdots, N\}$, respectively.
Let $g_{kn}$ denote the channel gain of user $k$ on subcarrier $n$, $\forall k \in \mathcal K, n \in \mathcal N$.
In our proposed TFS strategy, each subcarrier can be shared by multiple users with time sharing,
as shown in Fig. 1.

\begin{figure}
  \centering
  \includegraphics[width=3.2in]{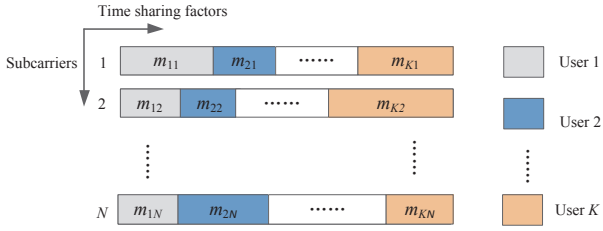}\\
  \caption{System model.}
\end{figure}

Let $m_{kn}\in[0,1]$ represent the time sharing factor of user $k$ on subcarrier $n$.
The achievable rate for information receiver of user $k$ can be expressed as
\begin{equation}\label{syseq1}
r_k=\sum_{n\in\mathcal N} m_{kn} B \log_2 \left(1+\frac{g_{kn}p_{kn}}{\sigma^2}\right),
\end{equation}
where $p_{kn}$ is the power allocated to user $k$ on subcarrier $n$, $\forall k \in\mathcal K, n \in \mathcal N$, $B$ is the bandwidth of each subcarrier, and $\sigma^2$ is the noise power.

Letting $\zeta \in(0,1]$ be the conversion efficiency of the energy harvesting process, the energy harvested by power receiver of user $k$ on subcarrier $n$ is $\zeta \sum_{l\in\mathcal K\setminus\{k\}} m_{ln} T p_{ln} g_{kn}$, where $T$ is the energy harvesting time.
Without loss of generality, we assume a normalized energy harvesting time $T=1$ in the following.
Hence, we can use both terms of energy and power interchangeably.
In the proposed TFS strategy, each user can independently either harvest energy or receive information among different subcarriers.
Thus, the total energy harvested by the power receiver of user $k$ can be written as
\begin{equation}\label{syseq2}
e_k=\zeta \sum_{l\in\mathcal K\setminus\{k\}}\sum_{n\in\mathcal N} m_{ln} p_{ln} g_{kn}.
\end{equation}

It should be noted that in the considered system a user can harvest energy from some subcarriers while it receives information from other subcarriers at the same time moment as in \cite{yinresource}.
Based on \cite{4623916},
transferring information and power separately on different subcarriers is
feasible and can be implemented by adapting band-pass filter based OFDMA
receiver.

In order to maximize the sum rate while guaranteeing the information and energy quality of service (QoS) constraints, we can optimize the power allocation strategy over multiple subcarriers.
The problem is formulated as:
\begin{subequations}\label{max1}
\begin{align}
\!\!\mathop{\max}_{\pmb{m}, \pmb{p}}\;
\!& \sum_{k\in\mathcal K}\sum_{n\in\mathcal N} m_{kn}B \log_2 \left(1+\frac{g_{kn}p_{kn}}{\sigma^2}\right)\\
\textrm{s.t.}\qquad \!\!\!\!\!\!\!\!\!
&
\sum_{n\in\mathcal N} m_{kn} B\log_2 \left(1+\frac{g_{kn}p_{kn}}{\sigma^2}\right)\geq R_k, \quad \forall k \in \mathcal K
\!\!\!\! \\
&\zeta \sum_{l\in\mathcal K\setminus\{k\}}\sum_{n\in\mathcal N} m_{ln} p_{ln} g_{kn} \geq E_k, \quad \forall k \in \mathcal K
\!\!\!\! \\
&\sum_{k\in \mathcal K}m_{kn}\leq 1, \quad \forall n \in \mathcal N\\
&\sum_{k\in \mathcal K}\sum_{n\in\mathcal N}m_{kn} p_{kn} \leq P_{\max}\\
&m_{kn}\geq 0, p_{kn}\geq0, \quad \forall k\in \mathcal K, n \in \mathcal N,
\end{align}
\end{subequations}
where $\pmb m=(m_{11}, \cdots, m_{K1}, \cdots, m_{KN})^T$, $\pmb p=(p_{11}, \cdots,$ $p_{K1}, \cdots, p_{KN})^T$, $R_k$ is the minimal transmission rate demand of user $k$, $E_k$ represents the minimal harvested energy of user $k$, and $P_{\max}$ stands for the maximal average power of the base station.
Constraints (\ref{max1}d) reflect the time sharing among different users, $m_{kn}$ is the allocated fraction of time for user $k$ on subcarrier $n$ to receive wireless information, while the rest time $\sum_{l\in \mathcal K\setminus\{k\}}m_{ln}$ is allocated to user $k$ to harvest energy on subcarrier $n$.
Note that the objective function of this paper is the sum rate maximization under the energy constraints of the users, which means the maximal rate of all users when the transmission energy of users merely comes from the harvested energy.

Assume that the time sharing factors of all the subcarriers are the same for each user, i.e., $m_{k1}=m_{k2}=\cdots=m_{kN}$, $\forall k \in \mathcal K$, problem (\ref{max1}) reduces to the rate maximization problem for TS strategy \cite{zhou2014wireless}.
Accordingly, TS  can be viewed as a special cases of the TFS strategy.
In \cite{zhou2014wireless}, there was no closed-form expression for optimal solution and the rate maximization problem for TS strategy was solved by iteratively optimizing time sharing factor and transmission power.
In the following, we transform problem (\ref{max1}) into an equivalent convex problem by an appropriate variable transformation, and obtain the semi closed-form expression of the optimal solution {\color{myc1}{via}} solving a modified problem of the original {\color{myc1}{rate maximization}} problem.

\section{Optimal Solution}

In this section, we first provide the optimal condition for problem (\ref{max1}), and then propose an iterative algorithm {\color{myc1}{based on semi closed form.}}

\subsection{Optimal Condition}
Problem (\ref{max1}) is nonconvex due to objective function (\ref{max1}a) and constraints (\ref{max1}b), (\ref{max1}c), (\ref{max1}e).
Thus, we first reformulate problem (\ref{max1}) by introducing a set of new non-negative variables:
$q_{kn} =m_{kn} p_{kn}$, $k=1, \cdots, K$, $n=1, \cdots, N$.
Then, problem (\ref{max1}) is equivalent to the following problem.
\begin{subequations}\label{max2}
\begin{align}
\!\!\mathop{\max}_{\pmb{m},{\pmb{q}}}\;
\!& \sum_{k\in\mathcal K}\sum_{n\in\mathcal N} m_{kn}C\ln \left(1+\frac{g_{kn}q_{kn}}{\sigma^2m_{kn}}\right)\\
\textrm{s.t.}\qquad \!\!\!\!\!\!\!\!\!
&
\sum_{n\in\mathcal N} m_{kn} C \ln \left(1+\frac{g_{kn}q_{kn}}{\sigma^2m_{kn}}\right)\geq R_k, \quad \forall k \in \mathcal K\\
&\zeta \sum_{l\in\mathcal K\setminus\{k\}} \sum_{n\in\mathcal N} q_{ln} g_{kn} \geq E_k, \quad \forall k \in \mathcal K
\!\!\!\! \tag{\ref{max2}c}\\
&\sum_{k\in \mathcal K}m_{kn}\leq 1, \quad \forall n \in \mathcal N \tag{\ref{max2}d}\\
&\sum_{k\in \mathcal K}\sum_{n\in\mathcal N}q_{kn} \leq P_{\max} \tag{\ref{max2}e}\\
&m_{kn}\geq 0, q_{kn}\geq0, \quad \forall k\in \mathcal K, n \in \mathcal N, \tag{\ref{max2}f}
\end{align}
\end{subequations}
where $\pmb q=(q_{11}, \cdots, q_{K1}, \cdots, q_{KN})^T$ and $C={B}/({\ln2})$.
Since
\begin{eqnarray*}
&&\!\!\!\!\!\!\lim_{m_{kn}\rightarrow0+}
m_{kn} C \ln \left(1+\frac{g_{kn}q_{kn}}{\sigma^2m_{kn}}\right)
\\
&&\!\!\!\!\!\!
=\lim_{x\rightarrow+\infty} \frac{C\ln \left(1+\frac{g_{kn}q_{kn}}{\sigma^2}x\right)}
{x}\\
&&\!\!\!\!\!\! =\lim_{x\rightarrow+\infty} \frac{C\ln \left(\frac{g_{kn}q_{kn}}{\sigma^2}\right)+C\ln x}
{x}
\\
&&\!\!\!\!\!\!=0,
\end{eqnarray*}
we define $m_{kn} C \ln \left(1+\frac{g_{kn}q_{kn}}{\sigma^2m_{kn}}\right)=0$ at $m_{kn}=0$.
From \cite[Page~89]{boyd2004convex}, the perspective function of $u(x)$ is the function $v(x,t)$ defined by $v(x,t)=tu(x/t)$, ${\textbf{dom}}\:v= \{(x,t)|x/t\in {\textbf{dom}}\:u, t>0\}$.
If $u(x)$ is a concave function, then so is its perspective function $v(x,t)$ \cite[Page~89]{boyd2004convex}.
{\color{myc1}{Because}} $\ln\left(1+\frac{g_{kn}q_{kn}}{\sigma^2}\right)$ is concave with respect to (w.r.t.) $q_{kn}$,
$m_{kn}\ln\left(1+\frac{g_{kn}q_{kn}}{\sigma^2 m_{kn}}\right)$ is concave w.r.t. ($m_{kn}, q_{kn}$).
{\color{myc1}{Due to the fact that}} (\ref{max2}a) is a nonnegative weighted sum of concave functions, the objective function of problem (\ref{max2}) is also concave w.r.t. ($\pmb{m}, {\pmb{q}}$) \cite[Page~79]{boyd2004convex}.
{\color{myc1}{Since}} the constraints of problem (\ref{max2}) are all convex, problem (\ref{max2}) is convex.
{\color{myc1}{As a result}}, we can obtain the following lemma.
\begin{lemma}
The optimal ($\pmb m^*, \pmb q^*$) of problem (\ref{max2}) satisfies $\sum_{k\in \mathcal K}m_{kn}^*= 1, \forall n \in \mathcal N$, and $\sum_{k\in \mathcal K}\sum_{n\in\mathcal N}q_{kn}^* = P_{\max}$.
\end{lemma}
\itshape \textbf{Proof:}  \upshape
Assume that the optimal solution of (\ref{max2}) is ($\pmb m^*, \pmb q^*$), where there exists at least one subcarrier $n$ with $\sum_{k\in \mathcal K}m_{kn}^*<1$.
Denoting $r_{kn}=m_{kn} C \ln \left(1+\frac{g_{kn}q_{kn}}{\sigma^2m_{kn}}\right)$, we have
\begin{equation}\label{firsetder1}
\frac{\partial r_{kn}}
{\partial m_{kn}}
=
C\ln\left(1+\frac{g_{kn}q_{kn}}{\sigma^2m_{kn}}\right)
-\frac{C g_{kn}q_{kn}}{{g_{kn}q_{kn}}+\sigma^2m_{kn}}.
\end{equation}
Define function $f(x)=\ln(1+x) -\frac{x}{1+x}$ for $x \geq 0$.
From (\ref{firsetder1}), we have $\frac{\partial r_{kn}}
{\partial m_{kn}}=C f(\frac{g_{kn}q_{kn}}{\sigma^2m_{kn}})$.
Since
\begin{equation}\label{functionf}
f'(x)=\frac{x}{(x+1)^2}> 0, \quad \forall x>0,
\end{equation}
we can obtain $f(x)> f(0)=0$, $\forall x>0$.
Then, $\frac{\partial r_{kn}}
{\partial m_{kn}} > 0$ and $r_{kn}$ is increasing for $m_{kn} > 0$.
Thus, the objective function (\ref{max2}a) can be further improved with an increment of $m_{kn}^*$, contradicting that the solution is optimal.

If the optimal $\pmb q^*$ of problem (\ref{max2}) {\color{myc1}{strictly satisfies constraint}} $\sum_{k\in \mathcal K}\sum_{n\in\mathcal N}q_{kn}^* < P_{\max}$,
the objective value (\ref{max2}a) can be improved with an {\color{myc1}{small}} increasing of power $q_{kn}^*$.
Hence, the conclusion is proved.
 \hfill $\Box$

\subsection{Semi Closed Form Based Efficient Solution}
{\color{myc1}{Because}} problem (\ref{max2}) is convex, {\color{myc1}{the popular
interior point method \cite{boyd2004convex,shadmand2010cross,bobarshad2009m,shikh2007joint,shikh2006apparatus,xu2011joint,chen2017caching,8379427,8382257,yang2018power}  can be used to obtain the globally optimal solution}}.
However, the complexity of solving problem (\ref{max2}) is $\mathcal O(K^3N^3)$ \cite[Page~561]{boyd2004convex} with the interior method from
the following Section $\text{\uppercase\expandafter{\romannumeral3}}$-C, which is in general not efficient.
Thus, we use the dual method with two steps to solve problem (\ref{max2}) with semi closed-form expression.

Assume that there exists one feasible solution ($\pmb m, \pmb q$) of problem (\ref{max2}) such that minimal rate constraints (\ref{max2}b) hold with inequality for at least one $k \in \mathcal K$.
This assumption is reasonable, as otherwise we can easily obtain the optimal value of problem (\ref{max2}) as $\sum_{k\in\mathcal K}R_k$, which is trivial.
Then, there exists $k\in\mathcal K$ such that $r_k>R_k>0$.
Without loss of generality, we assume that $m_{k1}>0$.
We reduce $m_{k1}$ to $m_{k1}'=m_{k1}-\epsilon$, where $\epsilon>0$ and
$\epsilon$ is set such that $r_k'>R_k$.
Then, we set $m_{l1}'=m_{l1}+\frac{\epsilon}{K-1}$, $\forall l \neq k$.
From the proof of Lemma 1, $r_{l}=\sum_{n\in\mathcal N}r_{ln}$ is increasing for $m_{l1}>0$ and
we can obtain $r_l'>r_l\geq R_l$, $\forall l \neq k$.
Thus, the Slater's condition is satisfied with new feasible solution ($\pmb m', \pmb q$) and the strong duality holds \cite[Page~265]{boyd2004convex}, which demonstrates that the dual method yields the optimal solution to the primal problem in (\ref{max2}).
Consequently, according to the dual theory \cite{bertsekas2009convex}, the optimal value of primal variables can be obtained by iteratively optimizing primal variables with given dual variables and updating dual variables with fixed primal variables.

In the first step, the primal variables are {\color{myc1}{optimized}} with {\color{myc1}{given}} dual variables.
To {\color{myc1}{optimize}} the primal variables, {\color{myc1}{we use the dual method}}.
The Lagrangian function of problem (\ref{max2}) is {\color{myc1}{given}} by
\begin{eqnarray*}
\mathcal L&&\!\!\!\!\!\!\!\!\!\!
(\pmb{m}, \pmb q, \pmb \alpha, \pmb \beta, \pmb\gamma, \lambda)
\!=\!
\sum_{k\in\mathcal K}\sum_{n\in\mathcal N} m_{kn} C \ln \left(1+\frac{g_{kn}q_{kn}}{\sigma^2m_{kn}}\right)
\\
 &&\!\!\!\!\!\!\!\!\!\!
+\sum_{k \in \mathcal K}\alpha_k\left(\sum_{n\in\mathcal N} m_{kn} C \ln \left(1+\frac{g_{kn}q_{kn}}{\sigma^2m_{kn}}\right)-R_k\right)
\\
 &&\!\!\!\!\!\!\!\!\!\!
+\sum_{k \in \mathcal K}\beta_k\left(\zeta \sum_{l\in\mathcal K\setminus\{k\}} \sum_{n\in\mathcal N} q_{ln} g_{kn}-E_k\right)
\\
 &&\!\!\!\!\!\!\!\!\!\!
+\!\!\sum_{n\in \mathcal N}\gamma_n\left(\!1\!-\!\sum_{k\in \mathcal K}m_{kn}\!\right)
\!+\!\lambda\left(\!P_{\max}\!-\!\!\sum_{k\in \mathcal K}\!\sum_{n\in\mathcal N}q_{kn}\!\!\right)
,
\end{eqnarray*}
where $\pmb \alpha=(\alpha_1, \cdots, \alpha_K)^T$, $\pmb \beta=(\beta_1, \cdots, \beta_K)^T$ and $\pmb \gamma=(\gamma_1, \cdots, \gamma_N)^T$.
$\pmb \alpha$, $\pmb\beta$, $\pmb \gamma$ and $\lambda$ are non-negative dual
variables associated with corresponding constraints of problem (\ref{max2}).
According to \cite[Page~267]{boyd2004convex}, the optimal solution should satisfy:
\begin{eqnarray}
\!\!\!\!\!\frac{\partial \mathcal L}{\partial m_{kn}}&&\!\!\!\!\!\!\!\!\!\!=
{(1+\alpha_k)C}{}\ln\left(1+
\frac{g_{kn}q_{kn}}{\sigma^2m_{kn}}\right)\nonumber
\\
&& \qquad
-\frac{(1+\alpha_k)C{g_{kn}q_{kn}}}{{\sigma^2m_{kn}}+{g_{kn}q_{kn}}}
-\gamma_n=0, \label{KKT1_1}\\
\!\!\!\frac{\partial \mathcal L}{\partial q_{kn}}&&\!\!\!\!\!\!\!\!\!\!\!=\!\frac{(1\!+\!\alpha_k)C{g_{kn}m_{kn}}}{{\sigma^2m_{kn}}\!+\!{g_{kn}q_{kn}}}
\!+\!\sum_{l\in\mathcal K\setminus\{k\}}\!\beta_l\zeta g_{ln}\!-\!\lambda\!=\!0.\label{KKT1_2}
\end{eqnarray}
Since the above two first-order equations are about the function of $\frac{q_{kn}}{m_{kn}}$, only $\frac{q_{kn}}{m_{kn}}$ can be displayed in closed form.
As a result, the values of $m_{kn}$ and $q_{kn}$ cannot be uniquely calculated from (\ref{KKT1_1}) and (\ref{KKT1_2}).
Heuristically, one can obtain the value of $m_{kn}$ and $q_{kn}$ by iteratively updating closed-form expression of $q_{kn}$ with fixed $m_{kn}$ and closed-form formulation of $m_{kn}$ with given $q_{kn}$.
By using this iterative mechanism, additional number of iterations is needed, which could increase the complexity of the algorithm.
In the following, we further characterize the semi closed-form expressions of $m_{kn}$ and $q_{kn}$
without additional number of iterations
by modifying the objective function (\ref{max2}a) of problem (\ref{max2}).

With a small positive constant $X$, the objective function (\ref{max2}a) is {\color{myc1}{modified}} as
\begin{equation}\label{newobj}
 \mathop{\max}_{\pmb{m},{\pmb{q}}}\;
\! \!\sum_{k\in\mathcal K}\!\!\sum_{n\in\mathcal N}\! m_{kn}C\ln \left(\!\!1\!+\!\frac{g_{kn}q_{kn}}{\sigma^2m_{kn}}\!\!\right)
\!\!+\!\!X\!\!\sum_{k\in\mathcal K}\!\!\sum_{n\in\mathcal N}\!\sqrt{m_{kn}}.
\\
\end{equation}
Since $X\sum_{k\in\mathcal K}\!\!\sum_{n\in\mathcal N}\!\sqrt{m_{kn}}\leq KNX$,
the upper-bound of the optimal-value gap between the modified optimization problem in (\ref{newobj}) with constraints (\ref{max2}b)-(\ref{max2}f) and the original problem in (\ref{max2}) is $KNX$.
Hence, the optimal solution of modified problem with objection function (\ref{newobj}) and constraints (\ref{max2}b)-(\ref{max2}f) {\color{myc1}{is approximately the same as}} the optimal solution of {\color{myc1}{original}} problem (\ref{max2}) if constant $X$ is sufficiently small.
There are two benefits of using square root in (\ref{newobj}).
The first benefit is that  square root is a concave function, which ensures that the modified optimization problem in (\ref{newobj}) with constraints (\ref{max2}b)-(\ref{max2}f) is a convex problem.
The second benefit is that the modified optimization problem in (\ref{newobj}) with constraints (\ref{max2}b)-(\ref{max2}f) yields semi closed-form expressions of $m_{kn}$ and $q_{kn}$.

With new objective function (\ref{newobj}), the optimal condition (\ref{KKT1_1}) should be modified as
\begin{eqnarray}
\!\!\!\!\!\frac{\partial \mathcal L}{\partial m_{kn}}&&\!\!\!\!\!\!\!\!\!\!=
{(1+\alpha_k)C} \ln\left(1+
\frac{g_{kn}q_{kn}}{\sigma^2m_{kn}}\right)\nonumber
\\
&&\!\!\!\!\!\!\!\!\!\!\!\!
-\frac{(1+\alpha_k)C{g_{kn}q_{kn}}}{{\sigma^2m_{kn}}+{g_{kn}q_{kn}}}
-\gamma_n+\frac{X}{2\sqrt{m_{kn}}}=0. \label{KKT2_1}
\end{eqnarray}
From (\ref{KKT1_2}), we can obtain
\begin{eqnarray}
q_{kn}
=\left[
\frac{(1+\alpha_k)C m_{kn}}
{
 \lambda-\sum_{l\in\mathcal K\setminus\{k\}}\beta_l\zeta g_{ln}
  }
-\frac{\sigma^2m_{kn}}{g_{kn}}
\right]^+,\label{KKT2_2}
\end{eqnarray}
where $[x]^+$ denotes $\max\{x,0\}$.
Substituting (\ref{KKT2_2}) into (\ref{KKT2_1}),
we are now able to obtain the unique solution to $m_{kn}$ as
\begin{equation}\label{KKT3_1}
m_{kn}
=\left.
\frac{X^2}
{\Xi^2}
\right|^1_0,
\end{equation}
 where $\Xi=-2
{(1+\alpha_k)C} \ln\left(
\frac{(1+\alpha_k)C{g_{kn}}}
{\sigma^2
\left(\lambda-\sum_{l\in\mathcal K\setminus\{k\}}\beta_l\zeta g_{ln}
\right)}
\right)
+2(1+\alpha_k)C
\left(1-
\frac
{\sigma^2
\left(\lambda-\sum_{l\in\mathcal K\setminus\{k\}}\beta_l\zeta g_{ln}
\right)}
{(1+\alpha_k)C{g_{kn}}}
\right)
+2\gamma_n$, $x|^1_0=\max\{0,\min\{x, 1\}\}$,
shown at the top of the next page.
Combing (\ref{KKT3_1}) and (\ref{KKT2_2}), we can easily have the unique value of $q_{kn}$ in closed form.

In the second step, we update the dual variables with the primal variables {\color{myc1}{optimized}} in the previous step.
{\color{myc1}{By exploiting}} the gradient based method in \cite{bertsekas2009convex},
the new values of the dual variables are updated by
\begin{eqnarray}
&&\!\!\!\!\!\!\!\!\!\!\!\!\!
\alpha_k{(t+1)}=
 \bigg[\alpha_k{(t)}+\theta(t) R_k\nonumber
\\
&& \quad
-
\theta(t) \sum_{n\in\mathcal N} m_{kn}(t)C \ln \left(\!1+\!\frac{g_{kn}q_{kn}(t)}{\sigma^2m_{kn}(t)}\!\right)
\bigg]^+,
\label{lagpara1}\\
&&\!\!\!\!\!\!\!\!\!\!\!\!\!
\beta_k(t+1)=\left[\beta_k(t)-
-\theta(t)\left(\zeta \sum_{l\in\mathcal K\setminus\{k\}} \sum_{n\in\mathcal N} q_{ln}(t) g_{kn} \right.\right.
\nonumber
\\
&&\qquad\qquad\qquad\qquad\qquad\qquad\qquad\:\:
-E_k\Bigg)\Bigg]^{+},
\label{lagpara2}
\\&&\!\!\!\!\!\!\!\!\!\!\!\!
\gamma_n(t+1)\!=\!\left[\!\gamma_n(t)\!-\! \theta(t)\left(\!1\!-\!\sum_{k\in \mathcal K}m_{kn}(t)\!\right)\!\right]^+\!\!,
\label{lagpara3}
\\&&\!\!\!\!\!\!\!\!\!\!\!\!
\lambda(t+1)\!=\!\left[\!\lambda(t)\!- \!\theta(t)\left(\!P_{\max}\!-\!\!\sum_{k\in \mathcal K}\!\sum_{n\in\mathcal N}q_{kn}(t)\!\!\right)\right]^+,
\label{lagpara4}
\end{eqnarray}
where $\theta(t)>0$ is a dynamically chosen stepsize.
A typical selection of $\theta(t)$ can be found in \cite[Page~295]{bertsekas2009convex}.
As a result, the semi closed form based (SCFB) algorithm to obtain the optimal solution of problem (\ref{max2}) is given in Algorithm 1.
\textcolor[rgb]{0.00,0.00,0.00}{
\begin{algorithm}[h]
\caption{: Semi Closed Form Based (SCFB) Algorithm}
\begin{algorithmic}[1]
\State Initialize $\pmb \alpha(0)$, $\pmb \beta(0)$, $\pmb \gamma(0)$, $\lambda(0)$, and the iteration number $t=0$.
\State With given $\pmb \alpha(t)$, $\pmb \beta(t)$, $\pmb \gamma(t)$, $\lambda(t)$, update $\pmb m(t)$ and $\pmb q(t)$ based on (\ref{KKT3_1}) and (\ref{KKT2_2}), respectively.
\State With given $\pmb m(t)$ and $\pmb q(t)$, update $\pmb \alpha(t+1)$, $\pmb \beta(t+1)$, $\pmb \gamma(t+1)$, $\lambda(t+1)$  based on (\ref{lagpara1}), (\ref{lagpara2}), (\ref{lagpara3}) and (\ref{lagpara4}), respectively.
\State
If the objective function (\ref{newobj}) converges, terminate.
Otherwise, set $t=t+1$ and go to step 2.
\end{algorithmic}
\end{algorithm}
}
\subsection{Complexity Analysis}
By exploiting the SCFB algorithm, sum rate maximization problem (\ref{max2}) can be effectively solved with globally optimal solution.
{\color{myc1}{By using the standard
interior point method, the complexity of solving problem (\ref{max2})  is $\mathcal O(K^3N^3)$ \cite[Page 487, 569]{boyd2004convex} due to the fact that the dimension of the variables in problem (\ref{max2}) is $2K N$.
}}
For SCFB, the {\color{myc1}{main}} complexity lies in
{\color{myc1}{obtaining}} time factor $\pmb m$ and power vector $\pmb q$.
To compute $m_{kn}$ by using (\ref{KKT3_1}), the complexity is $\mathcal O(K)$.
The complexity of solving $q_{kn}$ from (\ref{KKT3_1}) and (\ref{KKT2_2}) is $\mathcal O (K)$.
Thus, the total complexity of SCFB is
$\mathcal O(L
K^2N)$,
where $L$ is the total number of iterations of the proposed SCFB.
From Fig.~2 in Section IV, the value of total number of iterations is about 20.
Compared with the interior point method, we {\color{myc1}{observe}} that the proposed algorithm has a much lower order of complexity.

\subsection{Practical Implementation}
For the proposed TFS scheme, each user needs to upload the channel state information on each subcarrier to the base station via backhaul channel.
Based on the received channel state information, the base station performs the SCFB algorithm to obtain the optimal time sharing factor and power control.
The optimal optimal time sharing strategy is broadcasted to all users.
The base station simultaneously transfers wireless information and power according to the optimal power control strategy on different subcarriers,
and each user performs energy harvesting and information receiving on each subcarrier by adapting band-pass filter based OFDMA receiver \cite{4623916}.

\section{Numerical Results}
In this section, we evaluate the performance of the proposed TFS scheme.
The number of subcarriers is $N=15$.
The number of uses, $K$, is tested from 3 to 8.
The bandwidth of each subcarrier is $B=10$ $\text{MHz}$, and the noise power is $\sigma^2=-174$ $\text{dBm}/\text{Hz}$.
We set the maximal average power of the base station as $P_{\max}=17$ $\text{dBm}$.
The energy harvesting time $T=1$ s, and the conversion efficiency of the energy harvesting process is $\zeta=0.2$.
We assume equal minimal harvested power, i.e., $E_k=E$, $\forall k\in \mathcal K$.
The minimal rate requirement is 5 Mbps for each user.
Moreover, the path loss model is  $128.1+37.6\log_{10} d$ ($d$ is in km)
and the standard deviation of shadow fading
is $4$ dB \cite{access2010further}.

We first {\color{myc1}{investigate}} the convergence behavior of the proposed algorithm.
Fig. 2 illustrates the sum rate versus the number of iterations of the proposed algorithm under different values of parameter $X$.
It can be seen that the sum rate increases with the number of iterations and the convergent value with smaller $X$ is greater than with larger $X$.
Note that convergence number is pretty small compared to $KN^2=900$, which validates that the proposed algorithm has a lower complexity {\color{myc1}{than}} the interior method according to {\color{myc1}{Section}} $\text{\uppercase\expandafter{\romannumeral3}}$-C.
When $X=10^{-3}$, the upper-bound of the optimal-value gap between the modified optimization problem in (\ref{newobj}) with constraints (\ref{max2}b)-(\ref{max2}f) and the original problem in (\ref{max2}) is $KNX=0.06$.
Since the constant $X$ in (\ref{newobj}) affects the convergence speed and the optimal performance gap, the value $X=10^{-3}$ is carefully selected in the following simulations as a good trade-off between the convergence speed and the optimal performance gap.

\begin{figure}
  \centering
  \includegraphics[width=3.0in]{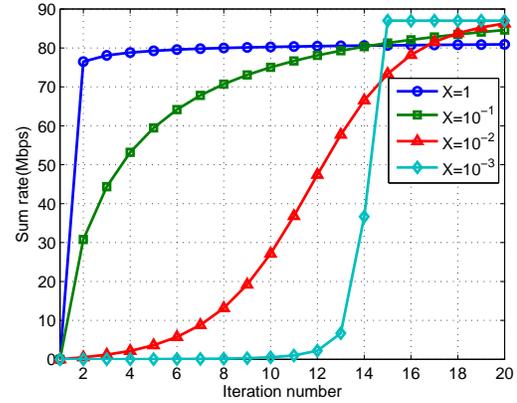}\\
  \caption{Convergence behavior under different values of parameter $X$ with $K=4$ and $E=36$ ${\mu}$W.}
\end{figure}

\begin{figure}
  \centering
  \includegraphics[width=3.0in]{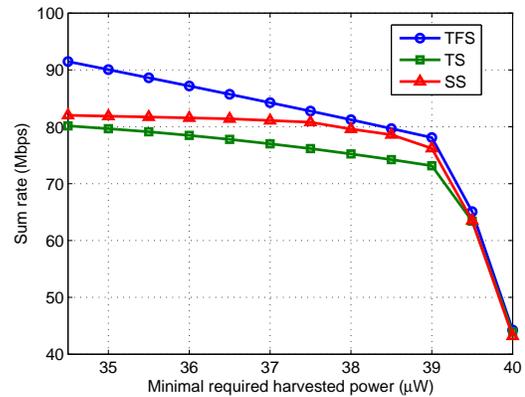}\\
  \caption{Sum rate versus minimal required harvested power with $K=4$.}
\end{figure}
We compare the proposed TFS strategy with the TS strategy proposed in \cite{zhou2014wireless}, where the received signal is either processed for energy harvesting or information decoding at each time,
and the suboptimal SS strategy proposed in \cite{yinresource}, where information and power are transferred separately on different subcarriers.
Fig.~3 {\color{myc1}{illustrates}} the sum rate comparison with different strategies of minimal required harvested power.
It is observed that the proposed TFS strategy outperforms the existing strategies, especially when the minimal harvested power is not too large.
This is because TS can be viewed as a special case of TFS and the feasible set of TFS is larger than that of TS, while the sum rate problem for SS is mixed integer programming problem and one practical suboptimal solution is provided \cite{yinresource}.
For all the three strategies, sum rate decreases with the minimum required harvested power, which is due to the fact that larger required harvested power means less power is left for transforming information.

The sum rate comparison with different numbers of users  is presented in Fig.~4.
It is observed that the sum rate increases as the number of users increases for all strategies, and the rate tends to be saturated for TS when the number of users is large.
This is because the increase of users can increase the total energy harvested by all users, which results in large transmission power and high sum rate.
It can be also found that the TFS is superior over TS and SS.

Computational complexity comparison is shown in Fig.~5, where ``TFS-SCFB'' refers to the time-frequency splitting strategy with the proposed semi closed form based algorithm, and ``TFS-interior point method'' refers to the time-frequency splitting strategy with the interior point method.
We can see that the TFS-SCFB has a lower complexity {\color{myc1}{than}} the TFS-interior point method.
The SS scheme yields the lowest complexity among the four algorithms, since only suboptimal solution is found \cite{yinresource}.
Fig.~6 illustrates the rate of each user.
Combing Fig.~3 and Fig.~6, it is interesting to observe that the proposed TFS  not only achieves the highest sum rate  but also yields best rate fairness among all comparing schemes.

\begin{figure}
  \centering
  \includegraphics[width=3.0in]{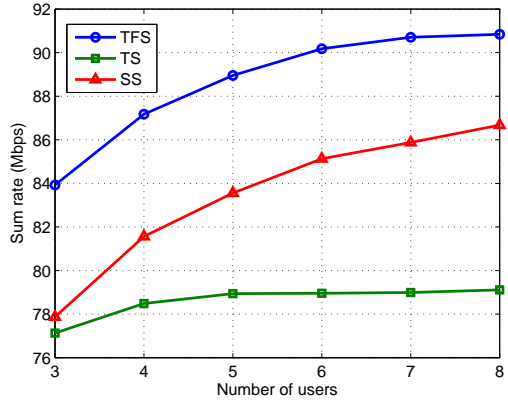}\\
  \caption{Sum rate versus number of users with $E=36$ ${\mu}$W.}
\end{figure}
\begin{figure}
  \centering
  \includegraphics[width=3.0in]{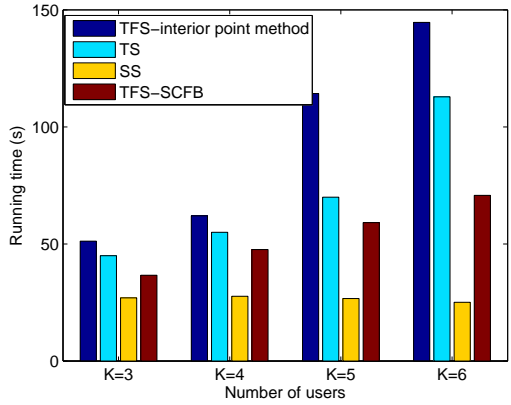}\\
  \caption{Running time versus number of users with $K=4$ and $E=36$ ${\mu}$W.}
\end{figure}

\begin{figure}
  \centering
  \includegraphics[width=3.0in]{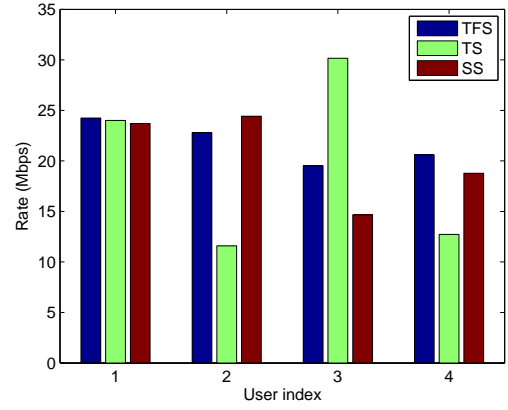}\\
  \caption{Rate of each user with $K=4$ and $E=36$${\mu}$W.}
\end{figure}

\section{Conclusion}
\vspace{-0.25em}
In this paper,
we propose a joint TFS strategy for multiuser SWIPT OFDM networks.
We investigate the sum rate maximization problem subject to both information and energy QoS constraints.
We first transform {\color{myc1}{the original nonconvex problem}} into an equivalent convex problem and then propose a {\color{myc1}{low-complexity}} algorithm to obtain the globally optimal solution.
Numerical results show that the proposed TFS strategy yields higher sum rate than the existing TS and SS strategies.

\vspace{-0.7em}
\bibliographystyle{IEEEtran}
\bibliography{IEEEabrv,MMM}

\end{document}